\documentclass[preprint]{revtex4-1}
\usepackage{epsf,epsfig}
\usepackage[psamsfonts]{amssymb}
\usepackage{bm}
\usepackage{graphicx}
\usepackage{dsfont}
\usepackage{amsmath}
\usepackage{natbib}
\usepackage{xcolor}
\usepackage{mathrsfs}

\newcommand{\be}{\begin{equation}}
\newcommand{\ee}{\end{equation}}
\newcommand{\ra}{\rangle}
\newcommand{\la}{\langle}
\newcommand{\bea}{\begin{eqnarray}}
\newcommand{\eea}{\end{eqnarray}}
\newcommand{\nn}{\nonumber}
\newcommand{\commentold}[1]{}
\begin{document}
%%%%%%%%%%%%%%%%%%%%%%%%%%%%%%%%%%%%%%%%%%%%%%%%%%%%%%%%%%%%%%%%%%%%%%%%%%%%%%%%%%%%%%%%%%%%%%%%
\title{Structured matter wave evolution in external time-dependent fields}
\author{Shohreh Janjan}
\email{sh.janjan@uok.ac.ir}
\affiliation{Department of Physics, University of Kurdistan, P.O.Box 66177-15175, Sanandaj, Iran}
\author{Fardin Kheirandish}
\email{f.kheirandish@uok.ac.ir}
\affiliation{Department of Physics, University of Kurdistan, P.O.Box 66177-15175, Sanandaj, Iran}
\date{\today}% It is always \today, today,
\begin{abstract}
\noindent In the present work, we have analyzed the motion of a structured matter wave in the presence of a constant magnetic field and under the influence of a time-dependent external force. We have introduced exact propagator kernels obtained from partial differential equations based on the Heisenberg equations of motion. The initial wave function is assumed as a Gauss-Hermite wave function. For the evolved wave function, we have obtained and discussed the uncertainties, orbital angular momentum, and the inertia tensor in the center of mass frame of the density function. From the point of view of the quantum interferometry of matter waves, and also non-relativistic quantum electron microscopy, the results obtained here are important and more reliable than the approximate methods like the axial approximation.
\end{abstract}
%
%\pacs{42.50.Ct, 42.50.Nn, 03.70.+k, 78.20.Ci, 78.67.Pt}
%
\keywords{}
\date{\today}
\maketitle
%
%%%%%%%%%%%%%%%%%%%%%%%%%%%%%%%%%%%%%%%%%%%%%%%%%%%%%%%%%%%%%%%%%%%%%%%%%%%%%%%%%%%%%%%%%%%%%%%%%%
\section{Introduction}\label{Introduction}
%%%%%%%%%%%%%%%%%%%%%%%%%%%%%%%%%%%%%%%%%%%%%%%%%%%%%%%%%%%%%%%%%%%%%%%%%%%%%%%%%%%%%%%%%%%%%%%%%%
\noindent The study of the behavior of charged particle beams is essential from both theoretical and experimental points of view. In the classical theory of charged particle beams, the purpose is to study the position and motion of the beam along the optical axis. In contrast, in the quantum theory of charged particles beams, one is interested in the time-evolution of the matter-wave or the density matrix describing the beam \cite{R1}. The study of the optical properties of quantum-charged particle beams has been a significant part of studies and research mainly devoted to the effect of the electromagnetic and gravitational fields on the quantum mechanical phase of the matter waves \cite{R2,R3,R4,R5}.
Quantum beams are directional beams consisting of a stream of charged particles like electrons, protons, ions, etc. moving almost in one direction. From wave-particle duality, this stream of particles can be interpreted as a matter wave propagating in the same direction \cite{R6}. The quantum dynamics of the matter wave is governed by a complicated many-body Schr\"{o}dinger equation containing both the external and internal electromagnetic fields \cite{R1}. In the scalar theory of quantum-charged beam optics, one deals with a stream of spin-$0$ charged particles \cite{R7,R8,R9}. A natural feature of all beams with the helical phase is the orbital angular momentum that can be generated in the laboratory \cite{R10}. Electron microscopy uses electron beams and wave-like properties of the electrons guided through magnetic lenses to achieve a clear and high-resolution image of the object. The free electron beams carrying orbital angular momentum (OAM) were investigated in a seminal paper by Bliokh \cite{R11}, which triggered extensive experimental works directed toward the generation and applications of structured electron beams \cite{R12,R13,R14}. For a non-interacting beam, one can consider the dynamics of a single particle or a matter wave.

Here, for a class of Hamiltonians describing the dynamics of a structured matter wave in the external forces, we find the exact quantum propagators as solutions of partial differential equations originating from Heisenberg equations of motion for positions and momenta. For a recent progress on partial differential equations, the interested reader is referred to \cite{pde1,pde2,pde3,pde4}. The initial state of the matter wave can be a Hermite–Gaussian mode which is commonly used to represent the modes of an optical resonator when they maintain approximately parabolic phase profiles. Laser resonant modes are often Hermite-Gaussian because they experience simple changes in their phase and intensity profiles. The Laguerre–Gauss modes may also be used as the initial state of the matter wave with rotational symmetry. Since lasers often have components that break rotational symmetry, Hermite-Gaussian modes are more suitable for laser beams. We consider the initial state of the matter wave as a Gauss-Hermite wave function. For the evolved wave function, we obtain the uncertainties, orbital angular momentum, and the inertia tensor in the center of mass frame of the density function. The paraxial approximation is used in various laser and optical fiber physics phenomena. It is valid as long as the divergence angles are very limited, and the beam radius at the beam waist must be much larger than the wavelength \cite{R15,R16,R17}. Also, the paraxial approximation is not applicable when the deviation from the propagation direction due to the effect of external forces is not negligible. Therefore, we follow an exact approach to find the propagators. The exact results obtained here can have applications in the interference of matter waves and non-relativistic quantum electron microscopy. Quantum electron microscopy is one of the applications of quantum charged particle beam under the influence of magnetic field.  In this device, the electron beams are affected by the magnetic field. This magnetic field changes the path of electron beams and the beams are used as imaging signals \cite{R18,R19,okamoto2022resilient}. Atom interferometry is the practice of manipulating both the translational and internal states of atoms in a coherent manner, and is a crucial experimental technique for utilizing matter waves. Interference has played a significant role in scientific discoveries, from the initial experiments that proved the wave-like behavior of light to the groundbreaking accomplishments in matter wave interferometry involving electrons, neutrons, atoms, and even large molecules. It has provided new understandings of the laws of nature and has been an effective tool for metrology \cite{senthilkumaran2012interferometry,bommareddi2014applications,schaff2014interferometry}.
%%%%%%%%%%%%%%%%%%%%%%%%%%%%%%%%%%%%%%%%%%%%%%%%%%%%%%%%%%%%%%%%%%%%%%%%%%%%%%%%%%%%%%%%%%%%%%%%%%
\section{Matter-wave interacting with a linear time-dependent potential}\label{tdlp}
%%%%%%%%%%%%%%%%%%%%%%%%%%%%%%%%%%%%%%%%%%%%%%%%%%%%%%%%%%%%%%%%%%%%%%%%%%%%%%%%%%%%%%%%%%%%%%%%%%
\noindent The Hamiltonian for a matter wave under the influence of a time-dependent force is
\be\label{1}
\hat{H}=\frac{\hat{p}_x^2+\hat{p}_y^2+\hat{p}_z^2}{2 m} + \mu(t)\,\hat{x},
\ee
where $\mu(t)\,\hat{x}$ is the external potential exerting a time-dependent force $-\mu (t)$ on the matter-wave along the $x$-axis. Hermit-Gaussian beams are a subset of stable laser states with symmetry along the propagation axis. The initial wave function ($\psi(x,y,z;0)$) is considered a Hermite-Gauss wave function. Another important class of initial wave functions are Laguerre-Gauss wave packets having orbital angular momentum, and their time-evolution can be obtained similarly. The initial wave function, as a wave packet, starts its motion from the origin with the initial momentum $\hbar k_0/m$ in the $z$-axis direction.
\be\label{2}
\psi(x,y,z;0)=(\frac{\alpha}{\pi})^{\frac{1}{4}}\,\sqrt{\frac{2}{\omega_0^2 \pi 2^{m+n}\,n!m!}}\,\exp(-\frac{x^2+y^2}{\omega_0^2})\,\exp(-\frac{\alpha}{2} z^2 + i k_0 z)\,H_n(\frac{\sqrt{2}}{\omega_0} x)\,H_m(\frac{\sqrt{2}}{\omega_0} y),
\ee
where $\omega_0$ is the beam waist and $H_n (x)$ is the Hermite polynomial of order $n$.
Here, we follow a non-relativistic approach to find the density evolution of the matter wave $|\psi(x,y,z:t)|^2$. In the following, In Sec.III, we also consider the effect of a constant magnetic field by modifying the Hamiltonian through the minimal coupling model and studying the density evolution. To find the quantum propagators, we make use of explicit expressions for position and momentum operators in Heisenberg picture and properties of partial differential equations. Therefore, our preferred picture is the Heisenberg picture \cite{R20}.   
\begin{figure}
\centering
\includegraphics[scale=0.4]
{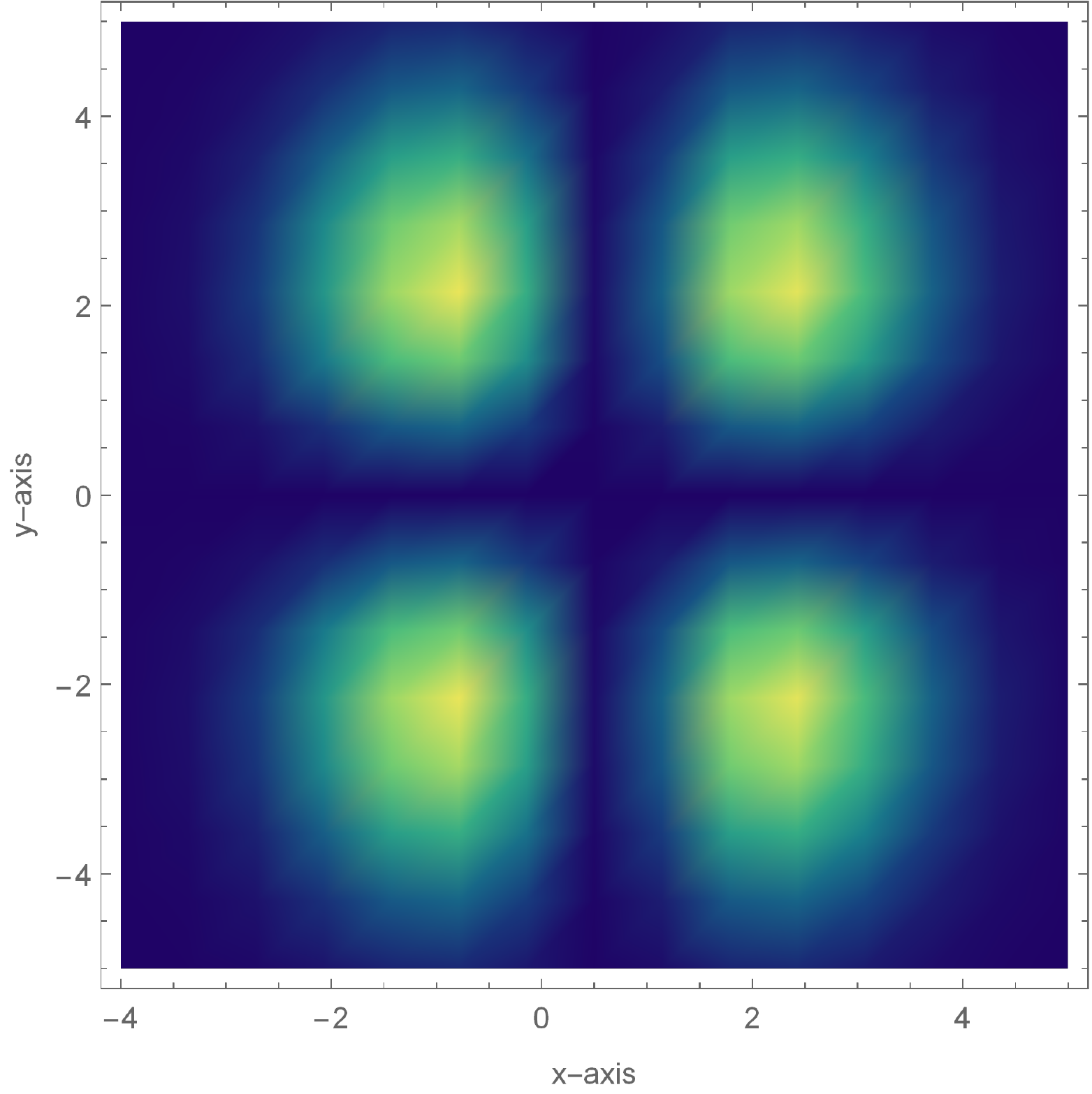}
\caption{(Color online) The density function $|\psi(x,y,z;0)|^2_{z=0}$ corresponding to Eq. (\ref{2}) on the x-y plane ($z=0$) for $n=m=1$.}
\end{figure}
By making use of the Hamiltonian Eq. (\ref{1}) and the Heisenberg equations for the momentum operators, we have $i\hbar\, \dot{\hat{p}}_x=[\hat{p}_x,\hat{H}]=-i\hbar\, \mu(t)$, $i\hbar\, \dot{\hat{p}}_y=[\hat{p}_y,\hat{H}]=0$, and $i\hbar\, \dot{\hat{p}}_z=[\hat{p}_z,\hat{H}]=0$. Therefore, the momentums along the $y$ and $z$ axis are constants of motion $\hat{p}_y(t)=\hat{p}_y(0),\,\,\, \hat{p}_z(t)=\hat{p}_z(0)$; any operator that is commuting with Hamiltonian is a constant of motion, so $\hat{p}_y$ and $\hat{p}_z$ are constants of motion but $\hat{p}_x$ evolves in time. The linear momentum along the $x$-direction is $\hat{p}_x(t)=\hat{p}_x(0)-\nu(t)$ where $\nu(t)=\int_0^t dt'\,\mu(t')$.
The Heisenberg equations for the position operators lead to
\bea\label{Heiseneqs}
&& \hat{x}(t)=\hat{x}(0)+(t/m)\hat{p}_x(0)-\xi(t)/m,\nn\\
&& \hat{y}(t)= \hat{y}(0)+(t/m)\hat{p}_y(0),\nn\\
&& \hat{z}(t)=\hat{z}(0)+(t/m)\hat{p}_z(0),
\eea
where we defined $\xi(t)=\int_0^t dt'\,\nu(t')$.
The classical trajectory of the center of mass of the matter wave can be found from the expectation values of the position operators in the initial state are $\langle\hat{x}(t)\rangle=\int dx\,dy\,dz\,\psi^*(x,y,z;0)\,\hat{x}(t)\,\psi(x,y,z;0)=-\xi(t)/m$, $\langle\hat{y}(t)\rangle=0$ and $\langle\hat{z}(t)\rangle=(\hbar k\,t)/m$. In a constant force field, we have $\xi(t)=\mu\,t^2/2$ and the trajectory will be a paraboloid as expected.

Orbital Angular Momentum (OAM) beams have a helical phase front along the axial center, which can be used for information transmission, imaging, and interferometry. The Orbital Angular Momentum (OAM) components corresponding to the matter wave are
\bea\label{OAMeqs}
&& \hat{l}_x(t)= \hat{y}(t)\,\hat{p}_z(t) -\hat{z}(t)\, \hat{p}_y(t)=[\hat{y}(0)+(t/m)\,\hat{p}_y(0)]\,\hat{p}_z(0)-[\hat{z}(0)+(t/m)\,\hat{p}_z(0)]\,\hat{p}_y(0)=\hat{l}_x(0),\nn\\  && \hat{l}_y(t)=[\hat{z}(0)+(t/m)\,\hat{p}_z(0)]\,[\hat{p}_x(0) -\nu(t)]-[\hat{x}(0)+(t/m)\,\hat{p}_x(0) -(\xi(t)/m)]\,\hat{p}_z(0),\nn\\
&& \hat{l}_z(t)=[\hat{x}(0)+(t/m)\,\hat{p}_x(0)-(\xi(t)/m)]\,\hat{p}_x(0)-[\hat{y}(0)+(t/m)\, \hat{p}_y(0)][\hat{p}_x(0)-\nu (t)].
\eea
The OAM along the $x$-axis is a constant of motion $\hat{l}_x(t)=\hat{l}_x(0)$. The expectation values of $\la\hat{l}_x(t)\ra$ and $\la\hat{l}_z(t)\ra$ in the initial state Eq. (\ref{2}) are zero, and for $\la\hat{l}_y(t)\ra$, one finds $\langle\hat{l}_y(t)\rangle=(\hbar k/m) \xi(t)+(\hbar k t/m) \nu(t).$

%%%%%%%%%%%%%%%%%%%%%%%%%%%%%%%%%%%%%%%%%%%%%%%%%%%%%%%%%%%%%%%%%%%%%%%%%%%%%%%%%%%%%%%%%%%%%%%%%%%%
\subsection{Exact propagator}
%%%%%%%%%%%%%%%%%%%%%%%%%%%%%%%%%%%%%%%%%%%%%%%%%%%%%%%%%%%%%%%%%%%%%%%%%%%%%%%%%%%%%%%%%%%%%%%%%%%%
\noindent Hamiltonian Eq. (\ref{1}) can be decomposed into commuting transverse ($\hat{H}_t$) and longitudinal ($\hat{H}_l$) parts as
\be\label{12}
\hat{H}=\underbrace{\frac{\hat{p}_y^2+\hat{p}_z^2}{2 m}}_{\hat{H}_{t}} +\underbrace{\frac{\hat{p}_x^2}{2m}+\mu(t)\,\hat{x}}_{\hat{H}_l},
\ee
therefore, the total quantum propagator can be written as
\be
k(x,y,z,t|x',y',z',t)=k_t(y,z,t|y',z',0)\,k_l(x;t|x',0),\nn
\ee
where $k_t (\cdot|\cdot) \,(k_l (\cdot|\cdot)) $ are transverse (longitudinal) propagators satisfying
\bea
&& \big[i\hbar\,\partial_t+\frac{\hbar^2}{2m}(\partial^2_y+\partial^2_z)\big]\,k_t(y,z,t|y',z',0)=0,\nn\\
&& \big[i\hbar\,\partial_t+\frac{\hbar^2}{2m}\partial^2_x-\mu(t)\,x\big]\,k_l(x,t|x',0)=0,\nn
\eea
with explicit solutions \cite{R20}
\bea\label{15}
k_t(y,z,t|y',z',0) &=& \Big(\frac{m}{2\pi i\hbar t}\Big)\exp\Big(\frac{im}{2\hbar t}[(y-y')^2+(z-z')^2]\Big),\nn\\
k_l(x,t|x',0) &=& \sqrt{\frac{m}{2\pi i\hbar t}}\,\exp\Big({\frac{-i}{2 m \hbar }\int_0^t dt'\,\Big(\frac{v(t')}{t'}\Big)^2}\Big)\nn\\
              &&\times\,\exp\Big(\frac{i m}{2 \hbar t}\Big[(x-x')^2-\frac{2\,v(t)}{m}(x-x')-\frac{2\,t\,\nu(t)\,x'}{m}\Big]\Big),\nn\\
\eea
where $v(t)=\int_{0}^{t} dt'\,t'\,\mu(t')$ and $\nu(t)=\int_{0}^{t} dt'\,\mu(t')$. Having the initial state Eq. (\ref{2}), we can find the evolved state $\psi(x,y,z,t)$ using
\be
\psi(x,y,z,t)=\int dx'dy' dz'\,k_{t}(y,z,t|y',z',0)k_l (x,t|x',0)\psi(x',y',z',0),\nn
\ee
leading to
\begin{multline}\label{pi0}
\psi(x,y,z;t)=\\
\Big(\frac{m\,\omega_0^2}{2}\Big)\Big(\frac{1}{i\hbar t\,(1+i\lambda)}\Big)\,(\frac{\alpha}{\pi})^\frac{1}{4}\,\frac{1}{\sqrt{2\pi\omega_0^2}}
\sqrt{\frac{m}{m+i\hbar \alpha t}}\, \Big(1-\frac{2}{1+i\lambda}\Big)^\frac{m}{2}\,\Big(1-\frac{2}{1+i\lambda}\Big)^\frac{n}{2}
\,\exp(-i\chi(t))\\ .\exp\Big(-\frac{k_0^2}{2\alpha}\Big) \exp\Big(-\frac{y^2}{\omega_0^2}\,(\frac{\lambda^2}{1+i\lambda}+i\lambda)\Big)\, \exp\Big(\frac{(ik_0-(im/\hbar t) z)^2}{2\alpha-2im/\hbar t}\Big)\, \exp\Big(-i\lambda(\frac{x^2}{\omega_0^2})-i\gamma\frac{\sqrt{2}\,x}{\omega_0}\Big)\\
.\exp\Big(-\frac{(\lambda \sqrt{2}\frac{x}{\omega_0} + \gamma + \frac{f}{\sqrt{2}})^2}{2+2i\lambda}\Big)\, H_m\left[\frac{y\,i\,\lambda \sqrt{2}}{\omega_0(1+i\lambda)\sqrt{1-\frac{2}{1+i\lambda}}}\right]\, H_n\left[\frac{i\,\lambda \sqrt{2}\frac{x}{\omega_0} + i\,\gamma + \frac{i\,f}{\sqrt{2}}}{(1+i\lambda)\sqrt{1-\frac{2}{1+i\lambda}}}\right],
\end{multline}
where for simplicity we have defined the time-dependent dimensionless functions $\lambda=-m\omega_0^2/2\hbar t$, $ \gamma=v(t)\,\omega_0/\sqrt{2}\,t\,\hbar$, $f=\omega_0 \,\nu(t)/\hbar$, and $\chi(t)=\frac{1}{2m\hbar}\int_0^t dt'\,(v(t')/t')^2$.
%
%
%%%%%%%%%%%%%%%%%%%%%%%%%%%%%%%%%%%%%%%%%%%%%%%%%%%%%%%%%%%%%%%%%%%%%%%%%%%%%%%%%%%%%%%%%%%%%%%%%%%%
\begin{figure}
\centering
\includegraphics[scale=0.4]
{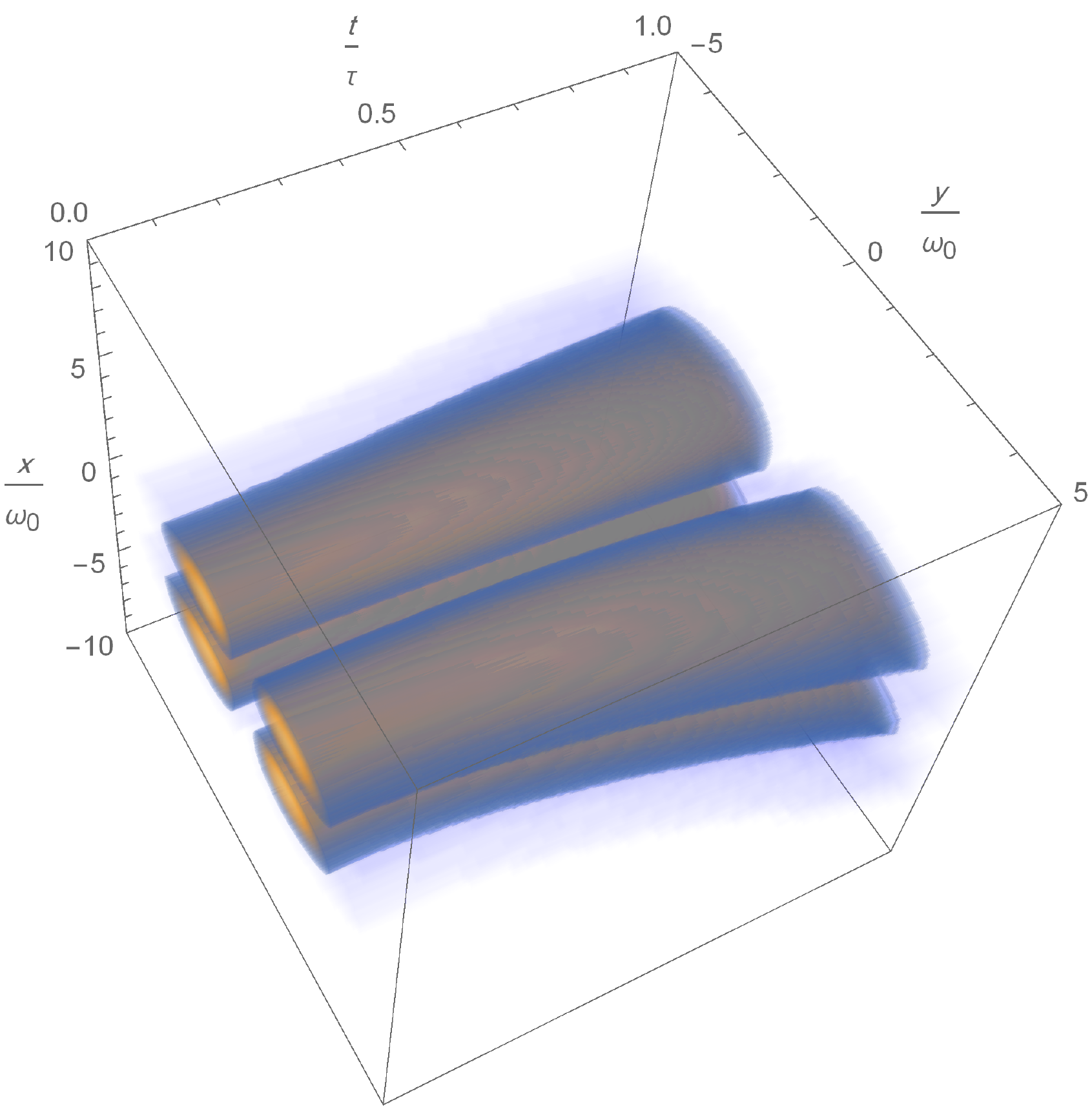}
\caption{(Color online) The density plot $|\psi(x,y,z;t)|^2_{z=0}$ of Eq. (\ref{pi0}), in terms of the scaled dimensionless variables $x/\omega_0$, $y/\omega_0$, $t/\tau$ where $\tau=m\omega_0^2/2\hbar$.}
\end{figure}

\begin{figure}
\centering
\includegraphics[scale=0.4]
{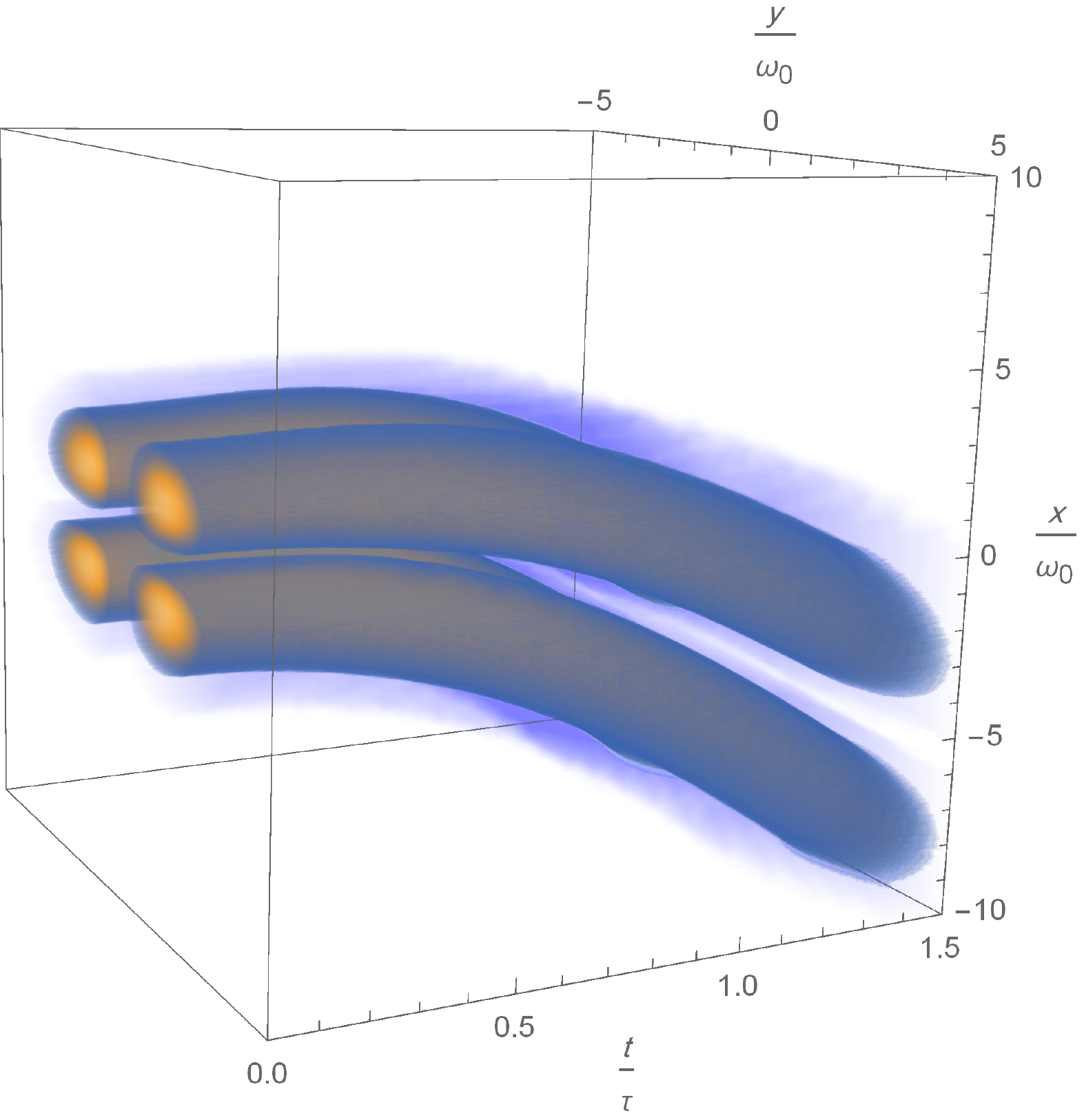}
\caption{(Color online) The density plot $|\psi(x,y,z;t)|^2_{z=0}$ of Eq. (\ref{pi0}) in the presence of a constant force. The scaled dimensionless variables are $x/\omega_0$, $y/\omega_0$, $t/\tau$ and $\tau=m\omega_0^2/2\hbar$.}
\end{figure}

\begin{figure}
\centering
\includegraphics[scale=0.4]
{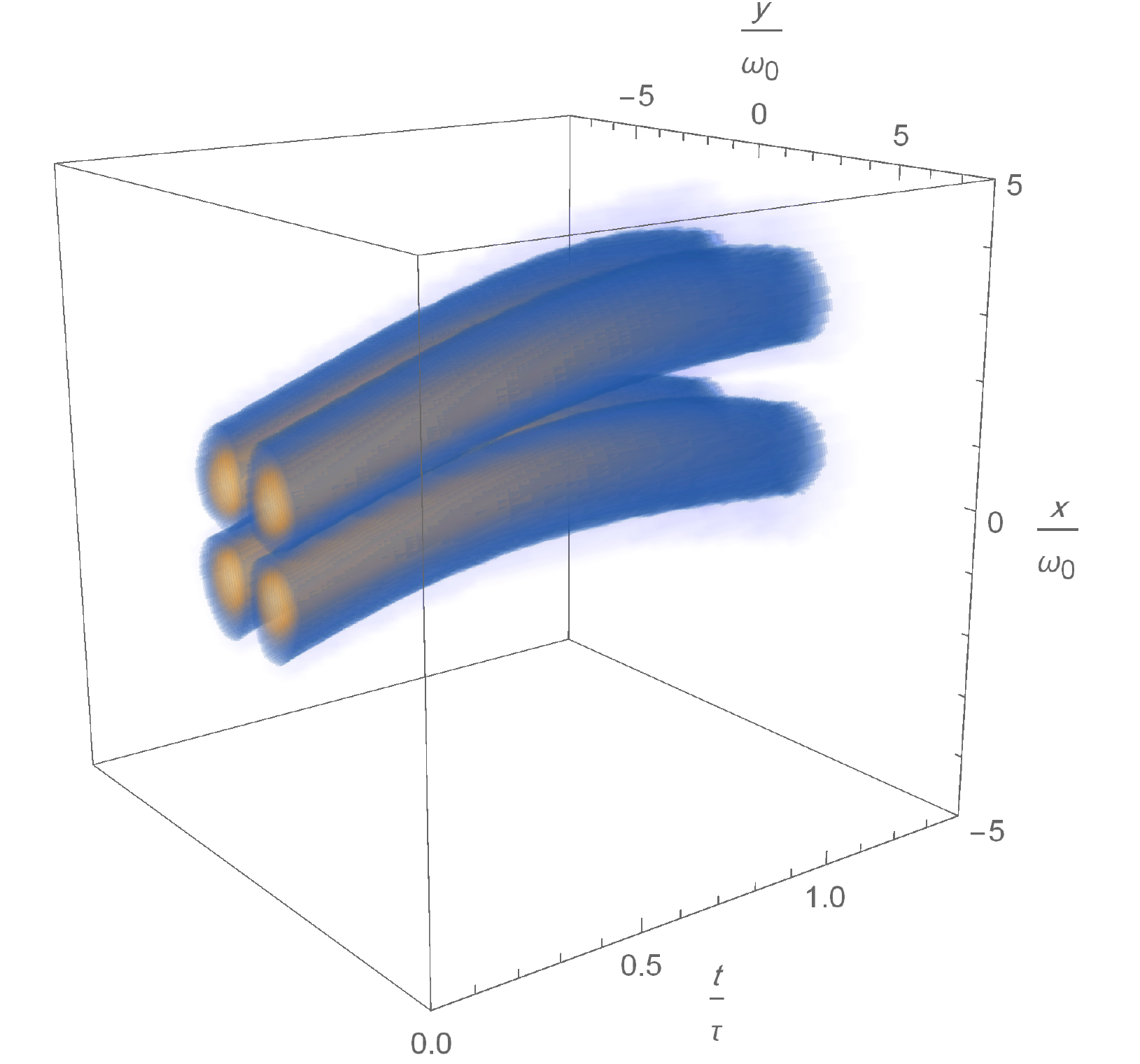}
\caption{(Color online) The density plot $|\psi(x,y,z;t)|^2_{z=0}$ of Eq. (\ref{pi0}) in the presence of the time-dependent force $\mu(t)=\mu_0\,\sin(2t/\tau)$ for $n=m=1$ and $\tau=m\omega_0^2/2\hbar$.}
\end{figure}

%%%%%%%%%%%%%%%%%%%%%%%%%%%%%%%%%%%%%%%%%%%%%%%%%%%%%%%%%%%%%%%%%
\subsection{Uncertainties and the Inertia Tensor }
%%%%%%%%%%%%%%%%%%%%%%%%%%%%%%%%%%%%%%%%%%%%%%%%%%%%%%%%%%%%%%%%%%
\noindent
By making use of Eqs. (\ref{Heiseneqs}), the expectation value of $\hat{x}(t)$ is $-\xi(t)/m$ and for $\hat{x}^2(t)$ we find
\bea
\la\hat{x}^2(t)\ra &=& \int dx\,dy\,dz\,\psi^*(x,y,z;0)\,\hat{x}^2(t)\,\psi(x,y,z;0),\nn\\
                   &=& (3\,\omega_0^2/4)\,+\,(3\,\hbar^2\,t^2)/(m^2\,\omega_0^2)\,+\,\xi^2(t)/m^2.
\eea
Therefore, the uncertainty in position is $\triangle x=(3 \omega_0^2/4+3 \hbar^2 t^2/m^2 \omega_0^2)^{1/2}$ which is spreading in time. Similarly, for $\la\hat{p}_x(t)\ra$ and $\la\hat{P}^2_x(t)\ra$ one easily finds
\bea
&& \la\hat{p}_x(t)\ra\,=\,\int dx\,dy\,dz\,\psi^*(x,y,z;0)\,[\-i\hbar \frac{\partial}{\partial x}-\nu(t)]\,\psi(x,y,z;0)\,=\,-\nu(t),\nn\\
&& \la\hat{p}^2_x(t)\ra\,=\,\int dx\,dy\,dz\,\psi^*(x,y,z;0)\,[\-i\hbar \frac{\partial^2}{\partial x^2}+\nu^2(t)]\,\psi(x,y,z;0)\,=\,(3\,\hbar^2/\omega_0^2)+\nu^2(t),\nn\\
\eea
with the variance
\be
\triangle p=\sqrt{\la\hat{p}^2_x(t)\Big\ra\,-\,\Big\la\hat{p}_x(t)\ra^2}=\frac{\sqrt{3}\,\hbar}{\omega_0}.
\ee
The moment of inertia along the $z$-axis is defined by
\bea
I_{zz}=m\la\hat{r}^2\ra &=& m\la\psi(0)|\hat{x}^2+\hat{y}^2|\psi(0)\ra,\nn\\
&=& \frac{3m}{2}\,\omega_0^2\,+\,\frac{3\,\hbar^2 t^2}{m\,\omega_0^2}\,+\,\frac{\xi^2(t)}{m},
\eea
and by using the general definition
\be
I_{ij}=m\la\hat{x}^2+\hat{y}^2+\hat{z}^2\ra_t\,\delta_{ij}-m\la\hat{x}_i\hat{x}_j\ra_t,
\ee
one can obtain the inertia tensor corresponding to the density function $|\psi(x,y,z;t)|^2$ as
\begin{align}
I_{ij}=\begin{bmatrix}
\varrho(t) & 0 & \frac{\hbar\,k\,t\,\xi(t)}{m}
\vspace{6mm}
\\
0 & \varrho(t)+\frac{\xi^2(t)}{m} & 0
\vspace{6mm}
\\
\frac{\hbar\,k\,t\,\xi(t)}{m} & 0 & \frac{3m^2\,\omega_0^4+12\hbar^2 t^2}{2m\omega_0^2}+\frac{\xi^2(t)}{m}
\end{bmatrix},
\end{align}
where for notational simplicity, we defined
\be
\varrho(t)=\frac{3m^2\,\omega_0^4+12\hbar^2 t^2}{4m\,\omega_0^2}+\frac{m^2+2\alpha\hbar^2 t^2(k^2+\frac{\alpha}{2})}{2m\alpha}.\nn
\ee
The inertia tensor with respect to the center of mass is defined by
\be
I_{ij}^{c}=m[\la\hat{x}^2+\hat{y}^2+\hat{z}^2\ra_t\,\delta_{ij}]+ m[R^{2}(t)\,\delta_{ij}-R_i\,R_j],\nn
\ee
where $R_i,\,(i=1,2,3)$ are the center of mass coordinates.
One can easily obtain the inertia tensor in the center of mass frame as
\begin{align}\label{23I}
I_{ij}^{c}=\begin{bmatrix}
\frac{3m}{4}\omega_0^2+\frac{3 \hbar^2 t^2}{m\,\omega_0^2}+\frac{m}{2\alpha}+\frac{\hbar^2 t^2 \alpha}{2m} & 0 & 0
\vspace{6mm}
\\0 & \frac{3m}{4}\omega_0^2+\frac{3 \hbar^2 t^2}{m\,\omega_0^2}+\frac{m}{2\alpha}+\frac{\hbar^2 t^2 \alpha}{2m} & 0
\vspace{6mm}
\\0 & 0 & \frac{3m}{2}\omega_0^2+\frac{6 \hbar^2 t^2}{m\,\omega_0^2}
\end{bmatrix}.
\end{align}
From Eq. (\ref{23I}) we deduce that during the evolution of the matter wave, the shape of the density function remains symmetrical ( $I_{xx} = I_{yy}$) with respect to the $x$ and $y$-axes in the center of mass frame though the volume of the density function is getting larger with respect to the center of mass axes. We also note that the components of $I_{ij}^{c}$ are independent on $\mu(t)$ and the density function just falls only in the direction of the external force.
%%%%%%%%%%%%%%%%%%%%%%%%%%%%%%%%%%%%%%%%%%%%%%%%%%%%%%%%%%%%%%%%%%%%%%%%%%%%%%%%%%%%%%%%%%%%%%%%%%%%%%%%%%%%%%%%%%%%%%%%%%%%%%%%%%%%%%%%%%%%%%%%%%%%%%%%%%%%%%%%%%%%%%%%%%%%%%%%%%%
\section{Propagation kernel in the presence of a constant magnetic field}
%%%%%%%%%%%%%%%%%%%%%%%%%%%%%%%%%%%%%%%%%%%%%%%%%%%%%%%%%%%%%%%%%%%%%%%%%%%%%%%%%%%%%%%%%%%%%%%%%%%%%%%%%%%%%%%%%%%%%%%%%%%%%%%%%%%%%%%%%%%%%%%%%%%%%%%%%%%%%%%%%%%%%%%%%%%%%%%%%%%
\noindent
Let us consider a charged particle with momentum $\mathbf{P}$ under the influence of a constant magnetic field in the direction of $z$-axis. The Hamiltonian is
\be
\hat{H}=\frac{(\mathbf{P}-q\,\mathbf{A})^2}{2 m}\,=\,\frac{\hat{p}_x^2}{2 m}\,+\,\frac{\hat{p}_y^2}{2 m}\,+\,\frac{\hat{p}_z^2}{2 m}\,+\,\frac{1}{2}m \omega^2 \hat{x}^2\,+\,\frac{1}{2}m \omega^2 \hat{y}^2\,+\, \omega \hat{p}_x\hat{y}\,-\,\omega  \hat{p}_y\hat{x},
\ee
where $\mathbf{P}=p_x \hat{i}\,+\,p_y \hat{j}\,+\,p_z \hat{k}$, $\omega=qB/2m$, and the vector potential is chosen as $\mathbf{A}=(-yB\hat{i}+xB\hat{j})/2$. From the Heisenberg equation of motion we find for the position and momentum operators
\bea\label{posi}
\hat{x}(t) &=& \alpha_R \hat{x}(0)-\alpha_I \hat{y}(0)\,+\,\frac{\beta_R}{m}\hat{p}_x(0)\,-\,\frac{\beta_I}{m}\hat{p}_y(0),\nn\\
\hat{p}_x(t) &=& -m\omega^2 \beta_R \hat{x}(0)+m\omega^2 \beta_I \hat{y}(0)+\alpha_R \hat{p}_x(0)-\alpha_I \hat{p}_y(0),\nn\\
\hat{y}(t) &=& \alpha_R \hat{y}(0)+\alpha_I \hat{x}(0)\,+\,\frac{\beta_I}{m}\hat{p}_x(0)\,+\,\frac{\beta_R}{m}\hat{p}_y(0),\nn\\
\hat{p}_y(t) &=& -m\omega^2 \beta_R \hat{y}(0)-m\omega^2 \beta_I \hat{x}(0)+\alpha_R \hat{p}_y(0)+\alpha_I \hat{p}_x(0),\nn\\
\hat{z}(t) &=& \frac{t}{m}\hat{p}_z(0)\,+\,\hat{z}(0),\nn\\
\hat{p}_z(t) &=& \hat{p}_z(0),
\eea
where $\alpha_R=\cos^2(\omega t)$, $\alpha_I\,=\,-\sin(\omega t)\cos(\omega t)$, $\beta_R\,=\,\cos(\omega t)\sin(\omega t)/\omega$ and $\beta_I\,=\,-\sin^2(\omega t)/\omega$. The kernel or Feynman propagator in the position space is defined as $\la\mathbf{r}|\hat{U}(t)|\mathbf{r'}\ra\,=\, K(\mathbf{r},t|\mathbf{r'},0)$ and can be determined using Eq. (\ref{posi}) as
\bea
K(\mathbf{r},t|\mathbf{r'},0)\,=&&\,\Big(\frac{m}{2 \pi i \hbar}\Big)^\frac{3}{2} \frac{\omega}{\sin(\omega t)\sqrt{t}}\exp\Big(\frac{i m}{2 \hbar t}(z-z')^2\Big)\nn\\
                                &&\times\,\exp\Big(\frac{-i m \omega}{\hbar}(x y'-x' y)\Big)\exp\Big(\frac{i m \omega \cot(\omega t)}{2 \hbar}(x-x')^2 +(y-y')^2\Big).\nn\\
\eea
Having the kernel, the time-evolution of the initial state
\be
\psi(x,y,z;0)=(\frac{\alpha}{\pi})^{\frac{1}{4}}\,\sqrt{\frac{2}{\omega_0^2 \pi 2^{m+n}\,n!m!}}\,\exp(-\frac{x^2+y^2}{\omega_0^2})\,\exp(-\frac{\alpha}{2} z^2 + i k_0 z)\,H_\textbf{n}(\frac{\sqrt{2}}{\omega_0} x)\,H_\textbf{m}(\frac{\sqrt{2}}{\omega_0} y),
\ee
is obtained from $\psi(\mathbf{r},t)=\int d^{3}\mathbf{r'}\,k(\mathbf{r},t|\mathbf{r'},0)\, \psi(\mathbf{r'},0)$ as
\begin{multline}\label{70}
\psi(x,y,z;t)=\left(\frac{ m}{2\pi i\hbar}\right)^\frac{3}{2}\left(\frac{\alpha}{\pi}\right)^\frac{1}{4}\frac{2\pi m\omega^2\left(y- x\,\cot(\omega t)\right)\left(\beta x+(m\omega/\hbar)y\,\cot(\omega t)\right)}{\hbar\,\omega_0^3\sqrt{2t}\sin(\omega t)\sqrt{\frac{\alpha}{2}-\frac{im}{2\hbar t}}\left(\frac{1}{\omega_0^2}-\frac{i(m\omega/\hbar)\cot(\omega t)}{2}\right)^3} \\
\times\,\exp\Big(\frac{imz^2}{2\hbar t}+\frac{(ik_0 - \frac{imz}{\hbar t})^2}{2\alpha-\frac{2im}{\hbar t}}\Big) \exp\Big(\frac{(i(m\omega/\hbar)y -i(m\omega/\hbar) \cot(\omega t)x)^2}{(\frac{4}{\omega_0^2} -2i(m\omega/\hbar)\cot(\omega t))}\Big)\\
\times\,\exp\Big(\frac{(-i(m\omega/\hbar) x -i(m\omega/\hbar)\cot(\omega t)y)^2}{(\frac{4}{\omega_0^2} -2i(m\omega/\hbar)\cot(\omega t))}\Big)\exp\Big(\frac{i(m\omega/\hbar)\cot(\omega t)}{2}(x^2+y^2)\Big).
\end{multline}
The probability density $\psi(x,y,z;t)\psi^{*}(x,y,z;t)$ has been depicted in FIG. 5. The applied magnetic field causes the particle to rotate in the direction of the field and gain orbital angular momentum.
\begin{figure}
\centering
\includegraphics[scale=0.4]
{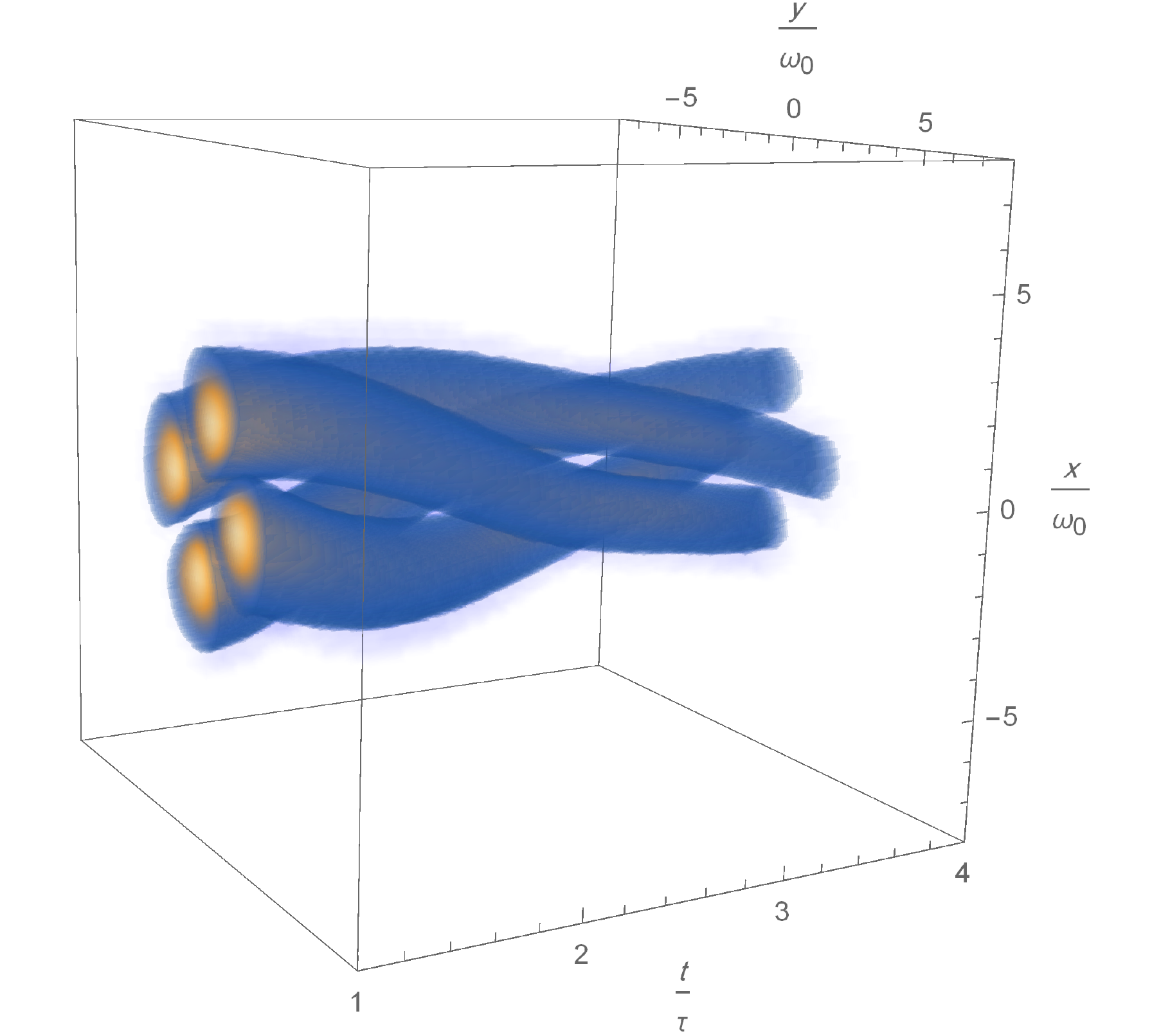}
\caption{(Color online) The density plot $|\psi(x,y,z;t)|^2_{z=0}$ of Eq. (\ref{70}), in terms of the scaled dimensionless variables $x/\omega_0$, $y/\omega_0$, $t/\tau$ where $\tau=m\omega_0^2/2\hbar$.}
\end{figure}
%%%%%%%%%%%%%%%%%%%%%%%%%%%%%%%%%%%%%%%%%%%%%%%%%%%%%%%%%%%%%%%%%%%%%%%%%%%%%%%%%%%%%%%%%%%%%%%%%%%%%%%%%%%%%%%%%%%%%%
\section{charged particle in the presence of a constant magnetic field under the influence of an external potential}
%%%%%%%%%%%%%%%%%%%%%%%%%%%%%%%%%%%%%%%%%%%%%%%%%%%%%%%%%%%%%%%%%%%%%%%%%%%%%%%%%%%%%%%%%%%%%%%%%%%%%%%%%%%%%%%%%%%%%%%
\noindent
In this section we generalize the problem investigated in the previous section by adding a linear potential $\mu\hat{x}$ to the Hamiltonian. The linear potential corresponds to a constant force like the gravitational force and we assume that the force is along the $x$-axis. Now the Hamiltonian is
\bea
\hat{H} &=& \frac{(\mathbf{P}-q\mathbf{A})^2}{2m}+\mu\hat{x},\nn\\
&=& \frac{\Big(\hat{p}_x + \frac{q B}{2}\hat{y}\Big)^2}{2m}+\frac{\Big(\hat{p}_y - \frac{q B}{2}\hat{x}\Big)^2}{2m} +\frac{\hat{p}_z^2}{2m}+\mu\hat{x},
\eea
where $\mu>0$ is the absolute value of the force along the $x$ axis.
Following the same process we did in the previous section, we find the position and momentum operators in Heisenberg picture as
\bea
\hat{x}(t) &=& \alpha_R\hat{x}(0)-\alpha_I\hat{y}(0)+\frac{1}{m}\beta_R\hat{p}_x(0)-\frac{1}{m}\beta_I\hat{p}_y(0)-\frac{\mu}{m}\xi_R,\nn\\
\hat{y}(t) &=& \alpha_R\hat{y}(0)+\alpha_I\hat{x}(0)+\frac{1}{m}\beta_I\hat{p}_x(0)+\frac{1}{m}\beta_R\hat{p}_y(0)-\frac{\mu}{m}\xi_I,\nn\\
\hat{p}_x(t) &=& \alpha_R\hat{p}_x(0)-\alpha_I\hat{p}_y(0)\,-\,m\omega^2\beta_R\hat{x}(0)\,+\,m\omega^2\beta_I\hat{y}(0)\,-\,\mu\eta_R,\nn\\
\hat{p}_y(t) &=& \alpha_R\hat{p}_y(0)+\alpha_I\hat{p}_x(0)\,-\,m\omega^2\beta_I\hat{x}(0)\,-\,m\omega^2\beta_R\hat{y}(0)\,-\,\mu\eta_I.
\eea
where for simplicity we have defined
\bea
&& \xi(t)\,=\,\int_{0}^{t}\beta(t')dt',\nn\\
&& \eta(t)\,=\,\int_{0}^{t} \alpha(t')dt',\nn\\
&& \alpha(t)\,=\,\alpha_R\,+\,i\alpha_I\,=\,\cos^2(t)\,-\,i\frac{\sin(2\omega t)}{2},\nn\\
&& \beta(t)\,=\,\frac{\sin(2\omega t)}{2\omega}\,-\,i\frac{\sin^2(\omega t)}{\omega}.
\eea
Having the explicit forms for position and momentum operators, the propagation kernel is obtained as
\begin{multline}
K(\mathbf{r},t|\mathbf{r'},0)=\Big(\frac{m}{2 \pi i \hbar}\Big)^\frac{3}{2} \frac{\omega}{\sin(\omega t)\sqrt{t}}\exp\Big(\frac{i m}{2 \hbar t}(z-z')^2\Big)\exp\Big(\frac{-i m \omega}{\hbar}(x y'-x' y)\Big)\\
.\exp\Big(\frac{i m \omega\cot(\omega t)}{2 \hbar}(x-x')^2 +(y-y')^2\Big)
\exp\Big(\frac{-i \mu t}{2\hbar}(x+x')\Big)\exp\Big(\frac{-i \mu}{2\hbar}(\frac{1}{\omega}-t\cot(\omega t)(y-y'))\Big),
\end{multline}
and the initial state Eq. (\ref{2}) evolves to
\begin{multline}\label{state}
\psi(x,y,z;t)=\left(\frac{ m}{2\pi i\hbar}\right)^\frac{3}{2} \left(\frac{\alpha}{\pi}\right)^{\frac{1}{4}}\,\exp\Big(\frac{(i(m\omega(y-x\cot(\omega t))-\frac{i\mu t}{2})^2}{(\frac{4\hbar^2}{\omega_0^2}-2i(m\hbar\omega)\cot(\omega t))}\Big)\\
\times\,\exp\Big(-\frac{i\mu t}{2\hbar}x\Big)\frac{2 \pi m\omega^2 \Big(y\omega-x\cot(\omega t)-\frac{\mu t}{2m\omega}\Big)\Big(\beta\hbar x+m\omega y\cot(\omega t)+\frac{\mu}{2}(\frac{1}{\omega}-t\cot(\omega t))\Big)}{\hbar^2\omega_0^3\sqrt{2t}\sin(\omega t)\sqrt{\frac{\alpha}{2}-\frac{im}{2\hbar t}}(\frac{1}{\omega_0^2}-\frac{i(m\omega/\hbar)\cot(\omega t)}{2})^3}\\
\times\,\exp\Big(\frac{(-im\omega (x+y\cot(\omega t)-\frac{i\mu}{2}(\frac{1}{\omega}-t\cot(\omega t))))^2}{(\frac{4\hbar^2}{\omega_0^2}-2im\hbar\omega\cot(\omega t))}\Big) \exp\Big(\frac{i\mu}{2\hbar}(\frac{1}{\omega}-t\cot(\omega t))y\Big)\\
\times\,\exp\Big(\frac{im\omega\cot(\omega t)}{2\hbar}(x^2+y^2)\Big)\,\exp\Big(\frac{imz^2}{2\hbar t}+\frac{(ik_0 - \frac{imz}{\hbar t})^2}{2\alpha-\frac{2im}{\hbar t}}\Big).
\end{multline}
The probability density of the state Eq. (\ref{state}) is depicted in FIG. 6. The applied magnetic field causes the particle to rotate in the direction of the field and gain an orbital angular momentum, and under the influence of external potential, the beam falls down in the $x$ direction.
%
%%%%%%%%%%%%%%%%%%%%%%%%%%%%%%%%%%%%%%%%%%%%%%%%%%%%%%%%%%%%%%%%%%%%%%%%%%%%%%%%%%%%%%%%%%%%%%%%%%%%%%%%%%%%%%%%%%%%%%%%%%%%%%%%%%%%%%%%%%%%%%%%%%%%%%%%%%%%%%%%%%%%%%%%%%%%%%%%%%%%%%%%%%%%%%%%%%%%%%%%%%%%%%%%%%%%%%%%%%%%%%%%%%%%%%%%%%%%%%%%%%%%%%%%%%%%%%%%
\begin{figure}
\centering
\includegraphics[scale=0.6]{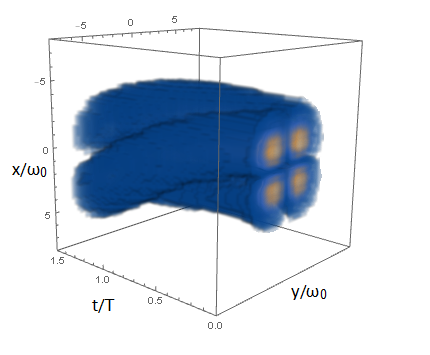}
\caption{(Color online) The density plot $|\psi(x,y,z;t)|^2_{z=0}$ of Eq. (\ref{70}), in the presence of a constant force in terms of the scaled dimensionless variables $x/\omega_0$, $y/\omega_0$, $t/\tau$ where $\tau=m\omega_0^2/2\hbar$.}
\end{figure}

For the initial Hermite-Gaussian wave function, the expectation value of $x$ was $\langle \hat{x}(t)\rangle=-\xi(t)/m$, the same result is obtained when the initial wave function is a Laguerre-Gauss state
\begin{multline}\label{LagurreGauss}
\psi_{LG}(\mathbf{r},0)= \sqrt{2p!/\pi w^2(z)(p+l)!}\, (\rho \sqrt{2}/w(z))^l\, \exp(-\rho^2/w^2(z))\, L^{l}_p(2\rho^2/w^2(z))\\
\times\, \exp(il\varphi)\, \exp(ik_0 \rho^2 z/2(z^2+z_{R}^2))\, \exp(-i(2p+l+1)\,tan^{-1}(z/z_R)),
\end{multline}
where $l$ is the azimuthal index, $p$ is the number of radial nodes and $w(z)$ is the beam radius \cite{R21}. For a constant external force $\mu(t)=\mu_0$, and in the absence of a magnetic field, both Hermite and Laguerre-Gauss wave functions lead to the same deviation $\langle \hat{x}(t)\rangle_{\mathbf{B}=0}=-\mu_0 t^2/2m$ from the propagation axis. By applying an external magnetic field and giving orbital angular momentum to the matter-wave, the expectation value of $x$ is $\langle \hat{x}(t)\rangle_{\mathbf{B}\neq0}=-\mu_0 \sin^2(\omega t)/2m\omega^2$, with a finite periodic deviation and accordingly more stable.
%%%%%%%%%%%%%%%%%%%%%%%%%%%%%%%%%%%%%%%%%%%%%%%%%%%%%%%%%%%%%%%%%%%%%%%%%%%%%%%%%%%%%%%%%%%%%%%%%%%%%%%%%%%%%%%%%%%%%%%%%%%%%%%%%%%%%%%%%%%%%%%%%%%%%%%%%%%%%%%%%%%%%%%%%%%%%%%%%%%%%%%%%%%%%%%%%%%%%%%%%%%%%%%%%%%%%%%%%%%%%%%%%%%%%%%%%%%%%%%%%%%%%%%%%%%%%%%
\section{Conclusions}
%%%%%%%%%%%%%%%%%%%%%%%%%%%%%%%%%%%%%%%%%%%%%%%%%%%%%%%%%%%%%%%%%%%%%%%%%%%%%%%%%%%%%%%%%%%%%%%%%%%%%%%%%%%%%%%%%%%%%%%%%%%%%%%%%%%%%%%%%%%%%%%%%%%%%%%%%%%%%%%%%%%%%%%%%%%%%%%%%%%%%%%%%%%%%%%%%%%%%%%%%%%%%%%%%%%%%%%%%%%%%%%%%%%%%%%%%%%%%%%%%%%%%%%%%%%%%%%
\noindent
For a class of Hamiltonians describing the propagation of a structured matter wave in the presence of a constant magnetic field and under the influence of a time-dependent external force, we found exact propagator kernels. The scheme was based on the solutions of partial differential equations originating from Heisenberg equations of motion and kernel properties. The initial wave function was a Gauss-Hermite wave function, and for the evolved wave function we studied the uncertainties, orbital angular momentum, and the inertia tensor in the center of mass frame of the density function. By exposing the matter-wave to an external magnetic field along the propagation direction, the structured matter wave gains an orbital angular momentum causing the entire structure of the matter-wave rotating around the propagation direction. The presence of a constant magnetic field brings the evolved state to a rather stable state by oscillating around a finite deviation $\sin^2(\omega t)/\omega^2$ compared to strictly increasing deviations in the absence of a magnetic field. The exact results obtained here can have applications in the non-relativistic quantum electron microscopy where the electron beams are affected by the magnetic fields and are used as imaging signals. Also, the results obtained here can be applied to atom interferometry and matter wave interferometry involving non relativistic electrons, neutrons, atoms, and even large molecules.
%%%%%%%%%%%%%%%%%%%%%%%%%%%%%%%%%%%%%%%%%%%%%%%%%%%%%%%%%%%%%%%%%%%%%%%%%%%%%%%%%%%%%%%%%%%%%%%%%%%%%%%%%%%%%%%%%%%%%%%%%%%%%%%%%%%%%%%%%%%%%%%%%%%%%%%%%%%%%%%%%%%%%%%%%%%%%%%%%%%%%%%%%%%%%%%%%%%%%%%%%%%%%%%%%%%%%%%%%%%%%%%%%%%%%%%%%%%%%%%%%%%%%%%%%%%%%%%%%

%\bibliographystyle{spmpsci}
\bibliography{myref}

\end{document}